\documentstyle[prl,aps,multicol,epsf]{revtex}
\tighten
\newcommand{\beq}{\begin{equation}}
\newcommand{\eeq}{\end{equation}}

\newcommand{\bea}{\begin{eqnarray}}
\newcommand{\eea}{\end{eqnarray}}

\begin{document}
\draft

\title{Lateral correlation of multivalent counterions is the
universal mechanism of charge inversion}
\author
{T. T. Nguyen, A. Yu. Grosberg, and B. I. Shklovskii}
\address{Department of Physics, University of Minnesota,
116 Church
St. Southeast, Minneapolis, Minnesota 55455} 

\address{\begin{quote}{\em We review works on screening of a macroion, such as colloidal 
particle or double helix DNA, by multivalent counterions. Multivalent metal
 ions, charged micelles, short or long polyelectrolytes can play the role of multivalent
 counterions. Due to the strong Coulomb repulsion such multivalent counterions form a 
strongly correlated liquid resembling a Wigner crystal at the surface of the macroion.
 Even if the macroion is neutralized by this liquid, a newly arriving counterion creates
in the liquid a correlation hole or image which attracts the ion to the surface.
 As a result, total 
charge of adsorbed counterions can substantially exceed the bare macroion charge, so that
 the macroion net charge inverts its sign. We discuss two previously suggested explanations of 
charge inversion and show that physics underlying both of them is intrinsically that 
of correlations, so that correlation is the universal force driving charge inversion.
}\end{quote}}

\maketitle

\begin{multicols}{2}

\section{Introduction}

Charge inversion is a counterintuitive phenomenon in which
a strongly charged particle (a
macroion) binds so many counterions
that its net charge
changes sign. As shown below the binding energy of a counterion
with large charge $Z$ is larger than
$k_B T$, so that this
net charge is easily observable: it is
the net charge that determines 
a particle drift in a weak field electrophoresis.
Charge inversion is possible for a variety of macroions, ranging
from the charged surface of mica 
to charged lipid membranes, colloids, DNA or actin.
Multivalent metal ions, small colloidal particles, charged micelles,
short or long polyelectrolytes including DNA can play the role of
multivalent counterions. 
Recently charge inversion has attracted significant attention
\cite{Roland,Linse,Perel99,Shklov99,Pincus,Bruinsma,Joanny00,Sens,Joanny1,Joanny2,Dubin,Lozada,Shklov001,Shklov0011,Holm,Shklov002,Stoll,Shklov003,Rubinstein,Andelman,Khokhlov,Matochiko}. 

Theoretically, charge inversion can be also thought of as an
over-screening.  Indeed, the simplest screening atmosphere,
familiar from the linear Debye-H\"{u}ckel (DH) theory, compensates at any
finite distance only a part of the macroion charge. It can be
proven that this is true also in the non-linear
Poisson-Boltzmann (PB) theory.
The statement that the net charge
preserves sign of the bare charge agrees with the common sense.
One can think that this statement is even more universal than
results of PB equation. 
However, this
presumption of common sense fails for screening by $Z$-valent
counterions with large $Z$ ($Z$-ions).
In this case, most of counterions are localized 
at the very surface of macroion.
The energy of their lateral Coulomb
interaction may exceed $k_BT$ by an order of
magnitude or more. 
As a result~\cite{Perel99,Shklov99}, at the macroion surface 
$Z$-ions form a two-dimensional (2D) strongly correlated
liquid (SCL) with properties resembling a Wigner crystal (WC), 
which is shown in Fig. 1.

The negative chemical potential of this
liquid leads to a correlation induced attraction
of $Z$-ions to the surface, which brings more of them
to the surface than necessary to neutralize it. 
Role of correlations of $Z$-ions  in another 
physical phenomenon attraction of two
likely-charged surfaces was recognized even earlier~\cite{Gulbrand}.
\begin{figure}
\epsfxsize=5cm \centerline{\epsfbox[0 -10 284 198]{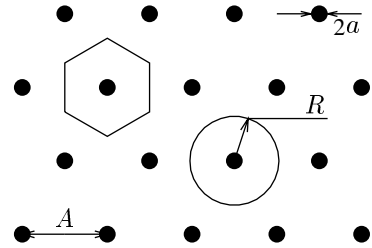}}
\caption{The Wigner crystal of positive $Z$-ions on
the negative uniform background of surface charge.
A hexagonal Wigner-Seitz cell and its simplified version as
a disk with radius $R$  are shown.}
\end{figure}
This paper consists of two parts. In the first part, 
we review recent works on correlation induced charge inversion.
Sec. II deals with small $Z$-ions, while in Sec. III the 
role of $Z$-ions is played by polyelectrolytes. In the second
part (Sec. IV), we discuss proposed 
alternative mechanisms of charge inversion: 
metallization approach (mental smearing of $Z$-ions on the macroion surface)
and counterion release. It is shown there that they are also based 
on the correlation physics, so that some of the
apparent difference is purely semantic. 

\section{Correlations of small size
multivalent counterions}

Let us consider screening a macroion surface
with a negative charge density $-\sigma$ by $Z:1$ salt with positive 
$Z$-ions in the case of large enough $Z$ and $\sigma$.
If such a surface is neutralized by $Z$-ions, 
the average distance between 
them in directions parallel to the plane equals
$R_0 = (\pi \sigma/Ze)^{-1/2}$. If this distance 
is much larger than the Gouy-Chapman length 
$\lambda = Dk_{B}T/ (2\pi Ze \sigma)$, which is the typical
distance to the surface mean field PB approach obviously fail.
The ratio of these two lengths $R_0/\lambda = 2\Gamma$, where 
$\Gamma = Z^{2}e^{2}/(R_0 D k_BT)$ is the dimensionless
inverse temperature measured in units of the
typical interaction energy and $D$ is the dielectric constant of water.
Here we are concerned with large enough $Z$ and $\sigma$ at which 
parameter $\Gamma \gg 1$. For example, 
at $Z=3$ and $\sigma = 1.0~e/$nm$^{2}$ we get
$\Gamma = 6.4$, $\lambda \simeq 0.08$ nm
and $R_0 \simeq 1.0$ nm. In such conditions Gouy-Chapman solution 
of PB equation can not be valid any more. Indeed,
if distance of a $Z$-ion to the surface $x$ is 
in the interval $\lambda \ll x \ll R_0$, it does not feel other $Z$-ions 
but interacts only with the macroion surface. This interaction leads 
to an exponential decay of concentration of $Z$-ions as function of $x$
in the range $ x \ll R_0$ instead of $1/x^2$ Gouy-Chapman law.
A new theory based on inequality $\Gamma \gg 1$ was
suggested in Ref.~\onlinecite{Perel99,Shklov99}.
The main idea of this theory is that at $\Gamma \gg 1$
the screening atmosphere can be divided in two distinct phases:
2D SCL, which should be treated exactly, and the gas phase
which at large distances $x$ can be desribed by the PB 
equation and is in equilibrium with 2D SCL.

Thermodynamic properties of 2D liquid of classical 
charged particles on the neutralizing background, 
so called one-component plasma (OCP) 
are well known (see bibliography in
Ref.~\onlinecite{Perel99,Shklov99}).
The chemical potential of OCP can be written as  
$\mu = \mu_{id}  + \mu_{c}$, where $\mu_{id}$ and $\mu_{c}$ are
its ideal and correlation parts. 
At $T=0$ OCP forms WC. To calculate $\mu_{c}$,
we start from the energy of WC per $Z$-ion,
\begin{equation}
\varepsilon(n) \simeq - 1.11 Z ^{2}e^{2}/RD
= - 1.96 n^{1/2}Z^{2}e^{2}/D,
\label{benergy}
\end{equation}
where $R=(\pi n)^{-1/2}$ is 
the radius of its Wigner-Seitz (WS) cell (See Fig. 1) and $n$
 is 2D concentration of $Z$-ions.
This gives $\mu_{c}= \mu_{WC}$, where
\begin{equation}
\mu_{WC} = \frac{\partial[n\varepsilon(n)]}{\partial n}
= - 1.65\Gamma k_BT = -1.65 {Z^{2}e^{2}}/{DR}. 
\label{muwc}
\end{equation}
 In the range of temperatures where
$1\ll \Gamma \ll 130 $ WC is melted and for SCL
$\mu_{c}= \mu_{WC} + \delta \mu$, where $\delta \mu$ is 
a small positive correction~\cite{Perel99,Shklov99}.

Physics of a large and negative $\mu_{WC}$
can be understood as follows.
Let us imagine for a moment that an insulating macroion is 
replaced by a neutral metallic particle. 
In this case, each $Z$-ion creates an
image charge of opposite sign inside the metal.
Energy of attraction to the image is $U(x) = -(Ze)^{2}/4Dx$,
where $x$ is the distance from $Z$-ion to the surface.
At the metal surface energy of the interaction 
with the image equals $U(a)= -(Ze)^{2}/4Da$, where 
$a$ is radius of the counterion. 
The chemical potential $\mu_{WC}$ for a charged surface
of an insulating macroion can be interpreted
in a similar language of images. 
\begin{figure}
\epsfxsize=6cm \centerline{\epsfbox[0 70 450 260]{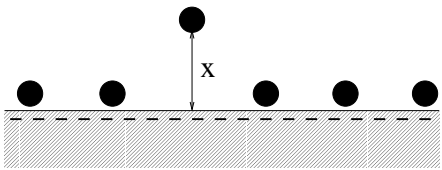}}
\vspace{0.2cm}
\caption{The origin of attraction of a new 
positive $Z$-ion to the already neutralized surface.
$Z$-ions are shown by solid circles. 
The new $Z$-ion creates its negative correlation hole.}
\end{figure}
Consider bringing a new $Z$-ion to the macroion surface already 
covered by an adsorbed layer of $Z$-ions (See Fig. 2). 
This layer plays 
the role of a metal surface. Indeed, the new $Z$-ion repels nearest
adsorbed ones, creating a correlation hole for himself. 
In other words, it creates a negative image.
Calculation of the energy of attraction to the image 
in this case is, however, less trivial than in the case 
of a metal. The problem is that minimal size of the 
image in the adsorbed layer is equal to the
WS cell radius $R$. (The adsorbed layer is a good metal 
only at larger scales.)
Thus, for WC we arrive at $\mu_{WC} \sim -(Ze)^{2}/DR$. 
Eq. (\ref{muwc}) provides the numerical coefficient in this expression.
It is clear now that contrary to the case 
of a metallic surface,
charge inversion for insulating macroion 
requires a finite density 
of an adsorbed layer, and, correspondingly, 
a finite bare charge density $\sigma$.

Correlation induced attraction of $Z$-ion to a neutral SCL
has two interesting analogies in the solid state
and atomic physics. The energy $|\mu_{WC}|$ 
plays the same role for $Z$-ions as the work function
plays for electrons of a metal or the ionization energy 
plays in a many-electron atom. It is known
that in the Thomas-Fermi
approximation which neglects exchange and correlation holes, 
both the work function of a metal~\cite{Lang} and the ionization energy 
of a many-electron atom~\cite{LLQM} vanish. For 
screening by $Z$-ions the PB 
approximation is an analog of the Thomas-Fermi approximation and
it results in $\mu_c = 0$. Only correlations lead to
a finite work function or ionization energy and to a 
finite $\mu_c \simeq \mu_{WC}$. 
  
Let us now discuss behavior of the 
concentration of $Z$-ions, $N(x)$
near the surface. For this purpose let 
us extract a $Z$-ion from SCL and move it 
along the $x$ axis. 
As it is mentioned above, this $Z$-ion leaves behind its
correlation hole. In the range
of distances $ x \ll R$, the correlation hole is
a disc of the surface charge
with radius $R$ (WS cell) and 
the $Z$-ion is attracted to the
surface by its uniform
electric field $E = 2 \pi \sigma/ D$. 
We can say that our $Z$-ion does 
not see other $Z$-ions until $x \ll R$.
Therefore, $N(x)= N_s \exp(-x/\lambda)$ at $x \ll R$.
Here $N_s = n/\lambda$ is the three-dimensional 
concentration of $Z$-ions at the plane. At large enough 
$x$ the concentration $N(x)$ should saturate at the level of
\begin{equation}
N_0 = N_s \exp(-|\mu_{WC}|/k_BT) = N_s \exp(-1.65\Gamma).
\label{N0}
\end{equation}
Let us find out how and where this saturation happens.
At $ x \sim R = 2\Gamma \lambda$ the ratio $x/\lambda$
reaches roughly speaking a half of $|\mu_{WC}|/k_B T$.
At $ x \gg R$ the $Z$-ion creates 
the image with relatively small radius $R$ which attracts 
$Z$-ion back and provides the decaying Coulomb correction to
the activation energy of $N(x)$:
$N(x)= N_s \exp\left(-[|\mu_{WC}| - Z^{2}e^{2}/4 Dx]/
k_B T\right)$. This correction is similar to the well known
"image" correction 
to the work function of a metal~\cite{Lang}. 
At $x = Z^{2}e^{2}/{4 D k_B T} = \lambda \Gamma^2/2$
this correction reaches $k_B T$,
so that $N(x)$ saturates at the value $N_0$.

The dramatic difference between the described above exponential 
decay of $N(x)$ and the Gouy-Chapman $1/(\lambda + x)^2$-law 
is obviously related to the correlation effects. 
Recently such an exponential decay of $N(x)$ 
with the following tendency to saturation
was rederived in more formal way and confirmed by Monte-Carlo 
simulations~\cite{Netz}.
At distances $x \gg \lambda \Gamma^2/2$,
interaction of the removed ion  
with its correlation
hole in SCL is not important and the correlation between
ions of the gas phase are even weaker
because $N(x)$ is exponentially small. Therefore, 
at larger distances one can describe $N(x)$ 
by PB equation~\cite{Perel99,Shklov99}. 

Studying charge inversion, we want to calculate  
the net charge density of the macroion surface
$\sigma^* = -\sigma + Zen >0$. Even though 
the system is not neutral, we can still 
use the chemical potential of a neutral OCP
given by Eq. (\ref{muwc}) after the following 
exact transformation. Indeed, let us  
add uniform charge densities $-\sigma^*$ 
and $\sigma^*$ to the macroion surface.
The first addition results in a neutral OCP.
The second addition creates only uniform potential
$e\psi(0)$ on the macroion 
surface. For example, if the macroion 
is a sphere with radius $r$
and screening radius of solution is larger than $r$ we get 
$\psi(0)=Q^*/Dr$,
where  $Q^*= 4\pi r^{2}\sigma^* = -Q + 4\pi r^{2} neZ $ is 
the net charge of the sphere and $-Q = 4\pi r^{2}\sigma$ is 
its bare charge. It is important to emphasize that
 macroscopic net charge $Q^*$ does not interact with 
OCP, because potential of the uniformly charged 
sphere at the neutral OCP is constant. 

The condition of balance of the electrochemical
 potential at the surface of 
macroion, $\mu$, and in the bulk of solution, 
$\mu_b$  now can be written as 
$Ze\psi(0) = - \mu_{WC} + (\mu_{b}- \mu_{id})=
 - \mu_{WC} + k_BT \ln(N_s / N) =k_BT \ln(N/N_0)$,
where $N$ is the concentration of $Z$-ions in the bulk of solution.
It is clear that when $N > N_0$, the net charge density
$\sigma^{*}$ is indeed positive, i.e. has the sign opposite
to the bare charge density $-\sigma$. The concentration $N_0$
is very small because $|\mu_{WC}|/k_BT =1.65\Gamma \gg 1$. Therefore,
it is easy to achieve charge inversion increasing $N$.
At large enough $N$ we have $|\mu_{b}- \mu_{id}| \ll |\mu_{WC}|$.
This gives a simple equation for calculation of
the maximal inverted charge density
\begin{equation}
\psi(0) = |\mu_{WC}|/Ze. 
\label{capacitor}
\end{equation}
Let us consider 
a sphere with charge $-Q$ screened by $Z:1$ salt
with a large concentration, $N$. In this case 
$\psi(0)=Q^*/Dr$ and Eq. (\ref{capacitor})
has a simple meaning: $|\mu_{WC}|/Ze$ is 
the "correlation" voltage which charges a spherical capacitor.
Expressing $R$ and $|\mu_{WC}|$ through $Q$ and
$Z$ we arrive at the simple prediction~\cite{Shklov99} 
for the maximum possible inverted charge:
\begin{equation}
Q^* = 0.83\sqrt{Q Z e}.
\label{Q}
\end{equation}
This charge is much larger than $Ze$, 
but still is smaller than $Q$ because of limitations 
imposed by the large charging energy of 
the macroscopic net charge. 
For example, for $Q=100e, Z=4$, we get $Q^* = 17e$. 
Eq. (\ref{Q}) was recently confirmed by numerical
simulations~\cite{Holm,Matochiko}.

It was also shown in Ref.~\onlinecite{Perel99,Shklov99}, that
in the case of a cylinder, the conventional picture
of nonlinear screening known as the Onsager-Manning (OM)
condensation should be strongly modified when
dealing with multivalent ions. Consider a cylinder with
a negative linear charge density $ - \eta$ and assume that 
$\eta > \eta_{z}$, where $\eta_{z} = k_BTD/Ze$.
OM theory~\cite{Manning}, shows that such 
a strongly charged cylinder is partially screened by 
counterions residing at 
its surface, so that net linear charge density
of the cylinder, $\eta^*= -\eta_{z}$.
 The rest of the charge is 
screened at much larger distances
according to the linear DH theory.
The OM theory uses PB mean field approach and,
therefore, does not take into 
account lateral correlations of counterions. 
It is shown in Ref.~\onlinecite{Perel99,Shklov99}
that the correlation induced negative 
chemical potential $\mu_{WC}$
leads to inversion of the sign of $\eta^*$ 
at $N > N_0$ 
in the case of cylinder, too.
At large enough $N$ inverted charge density 
$\eta^*$, can reach $k_BTD/e$.  

Even stronger charge inversion of a spherical or cylindrical
 macroion can be obtained 
in the presence of substantial concentration of 
monovalent salt, such as NaCl, in solution.
Monovalent salt screens long range Coulomb interactions stronger than 
short range lateral correlations between adsorbed $Z$-ions. 
Therefore, screening by monovalent salt diminishes the
charging energy of the macroion
much stronger than the correlation energy of $Z$-ions.
As a results, the inverted charge $Q^*$ becomes larger than 
that predicted by Eq. (\ref{Q}) and 
scales linearly with $Q$. 
Since, in the presence of a sufficient
concentration of salt,
the macroion is screened at the distance smaller 
than its size, the macroion can be thought of as an
over-screened surface, with inverted 
charge $Q^*$ proportional to the
surface area. In this sense, overall shape of the 
macroion and its surface is
irrelevant, at least to a first approximation.
Therefore, we
consider here a simple case: screening of a planar macroion
surface with a negative surface charge density $-\sigma$
by solution with concentration $N$ of $Z:1$ salt 
and a large concentration $N_1$ of a monovalent salt.
Correspondingly, we assume that all weak 
interactions are screened with
DH screening length
$r\!_s = \left(8\pi l_{B}N_1\right)^{-1/2}$.
For simplicity, we discuss here
 only a charged surface of insulating macroion
with the same dielectric constant as water.
Complications related to the difference 
between dielectric constants are discussed in
Ref.~\onlinecite{Shklov001}.

The dependence of the charge 
inversion ratio, $\sigma^{*}/\sigma$, on
$r\!_s$ was calculated 
analytically~\cite{Shklov001} in two limiting cases
$r\!_s\gg R_0$ and $r\!_s\ll R_0$, where 
$R_0 = (\pi \sigma/ Ze)^{-1/2}$ is the
radius a WS cell
at the neutral point $n=\sigma/Ze$. 
At $r\!_s\gg R_0$ calculation starts 
from Eq. (\ref{capacitor}).
Electrostatic potential can be calculated as 
potential of the plane with the charge density 
$\sigma^{*}$ screened at the distance $r\!_s$.
This gives $\psi(0) = 4\pi\sigma^{*}r\!_s$. 
At $r\!_s \gg R_0$ screening by monovalent ions
does not change Eq. (\ref{muwc}) substantially so that 
we still can use it in Eq. (\ref{capacitor}) which now
describes charging of a plane capacitor by voltage $|\mu_{WC}|/Ze$.
This gives~\cite{Shklov001}
\begin{equation}
\sigma^{*}/\sigma =  0.41 (R_0/r\!_s) \ll 1~~~~(r\!_s \gg R_0). 
\label{smallyI}
\end{equation}
Thus, at $r\!_s \gg R_0$ inverted charge
density grows with decreasing
$r\!_s$. Now we switch to the case of strong screening, $r\!_s \ll R_0$. 
It seems that in this case $\sigma^{*}$
should decrease with decreasing  $r\!_s$, because screening
reduces the energy of SCL and leads to its evaporation. In fact, this
is what eventually happens. However, there is
a range where $r\!_s \ll R_0$, but the energy of SCL is still large.
In this range, as $r\!_s$ decreases, the repulsion between
$Z$-ions becomes weaker 
and makes it easier to
pack more $Z$-ions on the plane.
Therefore, $\sigma^{*}$ continues to grow with
decreasing $r\!_s$. At $r\!_s \ll R_0$ it 
is convenient to minimize directly the free energy 
of WC of $Z$-ions at the macroin surface with respect of $n$.
This free energy consists of nearest neighbor
 repulsion energies of $Z$-ions 
and the attraction energy of $Z$-ions 
to the charge surface. All interactions are screened 
according to DH theory, so that interaction of non-nearest neighbors 
can be neglected. This gives~\cite{Shklov001}
\begin{equation}
F= -4\pi \sigma r\!_s Zen/D +
(3nZ^{2}e^{2}/DA)\exp(-A/r\!_s),
\label{SCWC}
\end{equation}
where $A = (2/\sqrt3)^{1/2}n^{-1/2}$
is the lattice constant of the hexagonal WC.
Minimizing this free energy with respect to $n$ one 
arrives at 
\begin{equation}
\sigma^{*}/\sigma
= (\pi/2\sqrt3) [R_{0}/r\!_{s}\ln(R_{0}/r\!_{s})]^{2}
~~~~(r\!_s\ll R_0).
\label{giantI}
\end{equation}
Thus $\sigma^{*}/\sigma$ grows with decreasing
$r\!_s$ and can become larger than 100\%.
At $r\!_s \sim R_0$ Eq. (\ref{giantI}) 
and Eq. (\ref{smallyI}) match each other.

\section{ Correlations of adsorbed polyelectrolyte molecules}

\subsection{Adsorption of rod-like polyelectrolytes on the charged
plane}

A practically important class of multivalent ions are
polyelectrolyte (PE) molecules.  In this section we discuss charge
inversion caused by adsorption of long rod-like $Z$-ions.  To make
signs consistent with the case of DNA, we assume
that PE charge is negative, $- \eta_0$ per unit length, while
the macroion surface is a plane with positive charge density $\sigma$.  The
basic idea is once again the strong lateral correlations: due to
the strong lateral repulsion, charged rods adsorbed at the surface
tend to be parallel to each other and have a short range order of
an one-dimensional WC in the perpendicular to rods direction (Fig.
3).
\begin{figure}
\epsfxsize=6.5cm \centerline{\epsfbox[50 540 310
660]{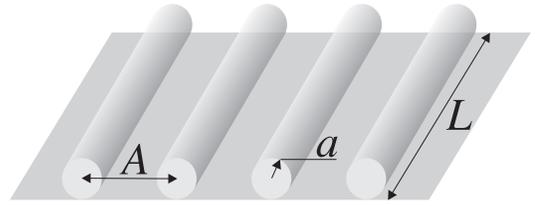}}
\caption{Rod-like negative $Z$-ions such as double helix DNA are
adsorbed on a positive uniformly charged plane.
Strong Coulomb repulsion of rods leads to
one-dimensional crystallization with lattice constant
$A$.}
\end{figure}
It is well known that the bare charge density of DNA, $-\eta_0$,
is about four times the critical density $-\eta_c =
k_BT/e$ of OM condensation. Therefore, about three quarters of
the bare charge of an isolated DNA is compensated by positive
monovalent ions residing at its surface so that the net charge of
DNA in the bulk solution is $\eta^* = -\eta_c$. This is generally not
true for adsorbed DNA.  Indeed, some OM-condensed
ions, as they are repelled from the charged surface, can be pushed
away from DNA and released into the solution when DNA is adsorbed
\cite{Bruinsma}. This effect is particularly important when
monovalent salt concentration is low, otherwise counterions do not
gain enough entropy to justify the release.

It was shown~\cite{Shklov001} that the criterion of counterion
release is governed by the comparison of screening length, $r\!_s$, 
of monovalent salt and the spacing between
critically charged rods, $A_0=\eta_c/\sigma$, when they
 neutralize the plane.  If screening is
strong and $r\!_s \ll A_0$, the potential of the surface is so
weak that counterions are not released.  In other words, charge
density for each adsorbed DNA helix remains $\eta^* = -\eta_c$.
Simultaneously, at $r\!_s \ll A_0$ the DH approximation can be
used to describe screening of the charged surface by monovalent
salt. Using these simplifications one can directly minimize the
free energy of one-dimensional crystal of DNA rods on the positive
surface written similarly to Eq. (\ref{SCWC}). Then the
competition between attraction of DNA helices to the surface and
the repulsion of the neighboring helices results in the negative
net surface charge density $-\sigma^*$.  The charge inversion
ratio reads~\cite{Joanny1,Shklov001} :
\begin{equation}
\sigma^{*}/\sigma = (\eta_c/ \sigma r\!_s)
/\ln(\eta_c/ \sigma r\!_s)~~~
(\eta_c/ \sigma r\!_s \gg 1, r\!_s \ll A_0).
\label{giant2}
\end{equation}
Thus the inversion ratio grows with decreasing $r\!_s$ as in the
case of spherical $Z$-ions. At small enough $r\!_s$ and
$\sigma$, the inversion ratio can reach 200\% before DNA molecules
are released from the surface. It is larger than for spherical
ions, because in this case, due to the large length of DNA helix,
the correlation energy remains large and WC-like short range order
is preserved at smaller values of $\sigma$ and/or $r\!_s$.  In the
works~\cite{Shklov001,Shklov0011}, we called this phenomenon "giant
charge inversion."

Let us switch now to the opposite extreme of weak screening by a
monovalent salt, $r\!_s \gg A_0$.  In this case, screening of the
overcharged plane by monovalent salt becomes strongly nonlinear,
with the Gouy-Chapman screening length $\lambda^* =Dk_{B}T/ (2\pi
e \sigma^*)$ much smaller than $r\!_s$.  Furthermore, some
counterions are released from DNA upon adsorbtion.  As a result
the absolute value of the net linear charge density of each
adsorbed DNA, $\eta^*$, becomes larger than $\eta_c$.  Two nonlinear
equations for $\sigma^*$ and $\eta^*$ are derived in
Ref.~\onlinecite{Shklov001} and are discussed below in Sec. IV.
 Their solution at $r\!_s \gg A_0$ reads:
\begin{equation}
\sigma^{*}/\sigma = (\eta_c/2\pi a \sigma) \exp \left( - \sqrt
{\ln (r\!_s/a) \ln (A_0/2\pi a)}\right), \label{NL1}
\end{equation}
\begin{equation}
\eta^*=  \eta_c  \sqrt {\ln (r\!_s/ a)/ \ln (A_0/2\pi a)} ~~~.
\label{NL2}
\end{equation}
At $r\!_s \simeq A_0 $ we get $\eta^* \simeq \eta_c$,
 $\lambda^* \simeq r\!_s$
and $\sigma^{*}/\sigma \simeq \eta_c/(2\pi r\!_s \sigma)$
 so that Eq.~(\ref{NL1})
crosses over smoothly to the strong screening  result of
Eq.~(\ref{giant2}). Since $\eta^*$ can not be smaller than $\eta_c$,
the fact that $\eta^* \simeq \eta_c$ already at $r\!_s \simeq A_0$
proves that at $r\!_s \ll A_0 $, indeed, $\eta^* \simeq \eta_c$.
Simultaneously, the fact that at this point $\lambda^* \simeq
r\!_s$ means that $r\!_s \ll \lambda^*$ at $r\!_s \ll A_0 $, so
that screening in this regime becomes linear and is described by 
DH theory.

\subsection{Flexible polyelectrolytes wrapping around the charged
spheres}

Until now we talked only about adsorption of a rod-like PE. If
the persistence
length $L_p$ is finite, PE is released from the charged surface
when $r\!_s$ is very small. This happens when interaction of a PE
segment of the length $L_p$ with the surface becomes smaller than
$k_BT$. On the other hand, at larger $r\!_s$ a flexible PE lays
flat at the surface. Due to its electrostatic rigidity neighboring
molecules form WC-like SCL and behave similarly to that of a
rod-like PE. The correlation induced adsorption of a flexible
weakly charged PE on an oppositely charged surface was
comprehensively studied in Ref.~\onlinecite{Rubinstein}. When
surface is covered by less than one complete layer of PE, results
are indeed close to presented above.

Let us also mention the charge inversion in the problem
of complexation of a positive sphere and a flexible negative PE (Fig.
4).
Refs.~\onlinecite{Pincus,Bruinsma,Sens,Joanny2} predicted substantial
charge inversion in this case:
more PE is wound around a sphere than necessary to neutralize it.
Role of correlations in such charge inversion was
recently emphasized in Ref.~\onlinecite{Shklov002}.
It was shown that neighbouring turns repel each
other and form almost equidistant solenoid,
which locally resembles WC. The tail of PE
repels PE adsorbed at the surface and creates a
correlation hole, which attracts the tail back to
the surface.
\begin{figure}
\epsfxsize=8cm \centerline{\epsfbox[0 0 330 140]{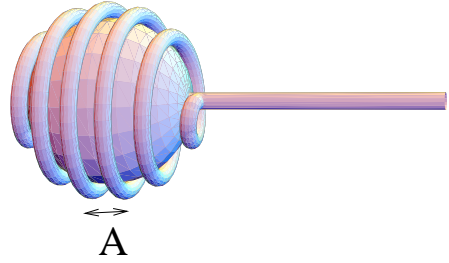}}
\caption{A PE molecule winding around a spherical macroion.
Due to Coulomb repulsion, neighboring turns lie parallel to each other.
Locally, they resemble a one-dimensional
Wigner crystal with the lattice
constant $A$.}
\end{figure}
The last example of the correlation driven charge inversion
we want to mention here is the complexation in a solution with
given concentrations of a long flexible negative PE and positive
spherical particles such
as colloids, micelles, globular proteins
 or dendromers~\cite{Shklov003}.
PE binds spheres winding around them.
If the total charge of PE in the solution is larger than
the total charge of spheres,
repulsive correlation of PE turns on a sphere surface
lead to inversion of the net charge
of each sphere. Negative spheres repel each other and form on
the PE molecule periodic beads-on-a-string structure,
which resembles 10 nm fiber of chromatin (Fig. 5).
\begin{figure}
\epsfxsize=9.0cm \centerline{\epsfbox{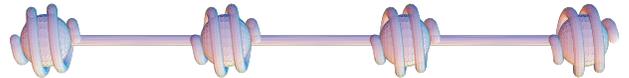}}
\caption{Complexation of a negative PE molecule
and many positive spheres in a necklace-like structure.
On the surface of a sphere neighboring
PE turns form WC similar to Fig. 4.
At larger scale, charged spheres repel each other and form
one-dimensional Wigner crystal along the PE molecule.}
\end{figure}
If the total charge of PE is smaller than
the total charge of spheres, the latter are under-screened
by PE and their net charges are positive.
Bound spheres once more repel each other
and form a periodic necklace.
Because a segment of PE wound around a sphere
interacts almost exclusively with this sphere,
it plays the role of WS cell. These
WC-like correlations lead to inversion of charge
of a PE molecule: spheres bind to PE in such a great number
that the net charge of the PE molecule becomes positive.
It is shown in Ref.~\onlinecite{Shklov003} that
inverted charge of PE by the absolute value can be
larger than the bare charge of PE even
in the limit of very weak screening by monovalent salt.
When a large concentration of a monovalent salt is added
charge inversion can reach giant proportions.
This theory is in a qualitative agreement with recent experiments
on micelles-PE systems~\cite{Dubin}.

\section{Are there other mechanisms of charge inversion?}

However briefly, we completed the review of works on correlation
induced charge inversion. It is now time to ask:  Are there
alternative, correlation-independent, electrostatic mechanisms
leading to this phenomenon?  Our answer is no, and we argue that
correlations-based mechanism is the universal one. Specifically,
we consider below two theories suggested in literature. One,
pioneered by \cite{Pincus,Joanny00}, 
considers adsorbed $Z$-ions as a smeared
continuous medium similar to a metal.  We argue that this {\em
metallization} approach is an approximation for correlation
theory; we discuss when this anzatz is accurate, and when it
fails. The other theory, put forward by \cite{Bruinsma}, views {\em
counterion release} as the driving force behind charge inversion.
To this end, we show that while counterion release is obviously
favorable for charge inversion, it is itself driven by
correlations.

\subsection{Metallization approach}

This approach was clearly formulated for the cases of complexation
of a PE with a sphere in \cite{Pincus} and for  adsorption of
flexible polymers on a charged plane in \cite{Joanny00}.  The
physical basis of this theory is a single, seemingly benign
approximation.  Namely, it considers the adsorbed $Z$-ions, at the
macroion surface, as a smeared continuum, while still treating
bulk solution as consisting of discrete charges.  Why is smearing
so much favorable as to cause adsorbtion of $Z$-ions to the
macroion which is already neutralized or even overcharged?

The answer is that smearing amounts to {\em complete} neglect of
self-energy $\varepsilon_{self}=Z^{2}e^{2}/2Da$ of adsorbed $Z$-ions; 
better to say, it neglects the
energy of each $Z$-ion's electric field in the
 range of distances between 
the ion radius $a$ and macroion size $r$. 
On the other hand, correlations can be
viewed as physical mechanism eliminating {\em a part} of
self-energy of each $Z$-ion corresponding to the field in 
the range of distances between correlation hole size
$R$ (which plays the role of screening radius of SCL)
and macroion size $r$. In this sense, metallization 
overestimates the role of correlations and the charge inversion
ratio, particularly when $R \gg a$.

To appreciate the difference between correlated and smeared set of
$Z$-ions, it is worth comparing both to the random distribution of the
same ions.  Correlated configuration is more favorable than
random: correlations happen because ions reconfigure themselves
non-randomly to gain some free energy.  Smeared continuum is also
more favorable than random configuration of discrete ions, the
difference being precisely the self-energy of ions.

Let us illustrate these ideas in the simplest geometry - a sphere
with charge $-Q$ and radius $r$ screened by spherical $Z$-ions of
radius $a \ll r$ in the absence of monovalent salt. Correlation
theory, as we have seen, results in Eq. (\ref{Q}) for this
problem.  What does metallization approach predicts for the same
problem? Assuming $a \ll r$, it is easy to write down all relevant
energies: $E_{met} = \left. Q^* \right.^2/2Dr$, which corresponds
to the uniformly smeared net charge $Q^*$; $E_{rand}= E_{met} + M
\varepsilon_{self}$,
which is the energy of the sphere with $M = 4\pi r^2 n$ randomly
distributed $Z$-ions on its surface;
and $E_{WC} = E_{rand} +
M\varepsilon(n)$, which is the energy of SCL of $Z$-ions. Here 
$\varepsilon(n)$ is the energy of WC per $Z$-ion given
 by Eq. (\ref{benergy}). The last
term of $E_{WC}$ is the correlation energy~\cite{LL}, it is the
negative (favorable) contribution due to the difference between
correlated and random configurations of $Z$-ions.  As expected,
$E_{met} < E_{WC} < E_{rand}$.

To address equilibrium charge inversion, let us now balance the
chemical potential of $Z$-ions at the surface and in the bulk.
Apart from translational entropy contribution negligible at high
enough concentrations of $Z$-ions, we have: for the bulk part chemical
potential is equal to the self-energy $\varepsilon_{self}$;  and
for the surface part $\mu = \partial E / \partial M$, where $E$ is
either $E_{met}$, or $E_{rand}$, or $E_{WC}$. Therefore,
depending on the approximation we want to use, the equilibrium
condition reads
\bea\label{balance} {\rm metal:} \ \ \ \varepsilon_{self} & = &
Q^{*}_{met} Ze/Dr \nonumber \\ {\rm random:} \ \ \
\varepsilon_{self} & = & Q^{*}_{rand} Ze/Dr +
\varepsilon_{self} \nonumber
\\ {\rm SCL:} \ \ \ \varepsilon_{self} & = & Q^{*}_{SCL}
Ze/Dr +
\varepsilon_{self} - 1.65 Z^2e^2/RD .
\eea
Looking at the result starting from the correlation
theory, we see that self-energies 
cancel and $Q^{*}_{SCL} = 1.65 Z e r /R$, producing Eq. (\ref{Q}).
By contrast, random distribution leads to $Q^*_{rand} = 0$, i.e.,
charge inversion is impossible without correlations.  Finally,
metallization approach yields $Q^{*}_{met} = Z e r /2a$. 
Comparing with result for SCL, metallization approach is off by a
numerical factor only when $R \sim 2a$, while it substantially
overestimates charge inversion at $R \gg a$.

We can repeat all the above arguments for adsorption of PE of
length $L < r$, radius $a$, and charge density below OM
threshold, $\eta < \eta_c$, on a sphere with radius $r$ and charge
$-Q$. At the surface of sphere PE molecules form one of several possible
anisotropic WC-like phases. For example, it can be nematic SCL
with distance $A$ between almost parallel neighboring PE molecules
or a solenoid with distance $A$ between turns, where PE molecules
continue each other simulating a long PE which winds around the
sphere.  In any case, balancing chemical potentials leads to the
equations similar to Eqs. (\ref{balance}):
\bea  {\rm metal:} \ \ \  \varepsilon_{self} & = & Q^{*}_{met}
\eta L/Dr  \nonumber
\\ {\rm random:} \ \ \   \varepsilon_{self} & = & Q^{*}_{rand} \eta L
/Dr + \varepsilon_{self} \nonumber
\\ {\rm SCL:} \ \ \  \varepsilon_{self} & = & Q^{*}_{SCL} \eta
L/Dr  + \varepsilon_{self} - (\eta^2/DL) \ln(L/A)  \ , \eea
where $\varepsilon_{self}=\eta^2 L D^{-1}\ln(L/a)$.  Thus, in the
correlation theory, we get $Q^*_{SCL}= (r \eta^2 L/Ze^2) \ln(L/A)$
because of cancellation of self-energies.  Random distribution of
positions and orientations of PE molecules does not lead to charge
inversion.  For metallized state, we have self-energy only in the
bulk, which gives $Q^*_{met} = (r \eta^2 L/Ze^2) \ln(L/a)$.  In
complete analogy with the previous case, we see that metallization
approach generally overestimates charge inversion, particularly at
$A \gg a$, i.e., when few PE molecules are adsorbed.  
With growing coverage
the metallization approximation becomes increasingly
accurate.

Qualitatively new effect, however, becomes important when macroion
is so strongly charged that its neutralization requires almost
full layer of $Z$-ions.  Excluded volume effect of hard cores of
$Z$-ions adds a positive term to the chemical potential of SCL,
which is proportional to $k_BT$ and diverges at the 
full coverage simultaneously with pressure.
Close to the full layer this term compensates and then
over-compensates the negative Coulomb term $\mu_{WC}$, so that
charge inversion disappears.  In the language of images, this
happens because a full layer is incompressible and does not allow
for correlation hole and image. For even larger macroion charge,
the second layer starts to form, launching a new wave of charge inversion.
In the beginning, charge inversion is small because all the attraction of a
new $Z$-ion approaching the surface is provided by a
weak interaction with an inflated image in the emerging second
layer, where once again $A \gg a$.  Continuing, we arrive at the
prediction of oscillating inverted charge $Q^{*}$ as a function of
$Q$ (see Fig. 6), where charge inversion vanishes every time when
the top layer of $Z$-ions is full~\cite{Shklov004}.  Metallization
approximation fails to capture the physics of these oscillations.
\begin{figure}
\epsfxsize=14cm \centerline{\epsfbox[0 530 475 640]{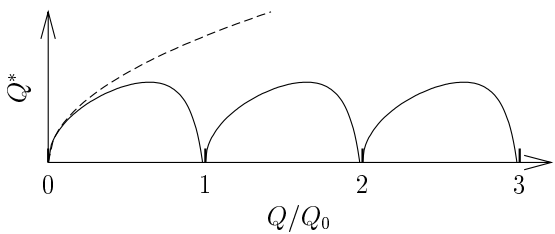}}
\caption{Schematic plot of the oscillating inverted charge of the
sphere $Q^*$ as a function of the absolute value $Q$ of its bare
charge. Here $Q_0$ is the charge of the full layer of
$Z$-ions. Dashed line obeys Eq. (\ref{Q}), which is derived for
$Z$-ions of vanishing radius $a$.}
\end{figure}

The oscillations of charge inversion we arrived at are similar to
the oscillations of the compressibility and magneto-capacitance in
the quantum Hall effect, which are related to consecutive filling
of Landau levels~\cite{Krav}. In this sense, we deal with a
classical analog of the quantum Hall effect.

We would like to conclude this subsection emphasizing the difference between
metallization approach and the PB one. PB approximation
effectively smears $Z$-ions everywhere, both at the macroion
surface and in the bulk of solution, while the metallization
approach keeps $Z$-ions of the bulk discrete.
 Therefore, it is not surprising that the metallization
approach somewhat overestimates charge inversion, while there is
no charge inversion in PB approximation.

\subsection{Counterion release}

Consider now adsorbtion of DNA, or other polymer charged above the
OM threshold $\eta_c$.  The result of correlation theory for
this case is presented in Sec. III.  Another, seemingly
independent reason for charge inversion, 
was suggested in \cite{Bruinsma}, arguing that
the entropy gain of counterions released by DNA {\it drives}
charge inversion.  Of course, this theory does not pretend for
universality, as it applies to highly charged polymers only.
Nevertheless, we want to understand the relations of counterion
release and correlations theories.

Let us start with a simple qualitative argument.  Imagine that
neutralizing DNA molecules, along with their OM-condensed
small ions, are adsorbing on the macroion (possibly releasing some
of the ions), and currently the neutralization condition is
achieved. Assume further that DNA rods are distributed randomly,
uncorrelated in both positions and orientations.  In this case,
next arriving DNA molecule feels no average field, so that it has
no reason to release its counterions. The situation is
completely different if DNA molecules are correlated on the
surface, where locally each molecule is surrounded by a
correlation hole - positive stripe of the background charge (WS
cell). The corresponding field, or positive potential of the WS
cell, may cause release of counterions not only at the neutrality
point, but even if the surface overall is overcharged ($\psi(0)<
0$). In other words, correlation hole, or adjustment 
of DNA molecules to each
other, or image charge, or correlations (all synonyms!) is a
necessary condition for {\em both} counterion release {\em and}
charge inversion.

To understand this better, let us re-examine the fundamental
physical conditions governing the adsorbtion equilibrium, namely
that chemical potentials of both DNA and small ions should be the 
same at the surface of macroion and in the bulk of the solution.   
The equations expressing these conditions were derived in
\cite{Shklov0011} (Eqs. (57) and (60) of that work). We re-derive
them below, as Eq. (\ref{first}) and
Eq. (\ref{DNAmu}). We deal below only 
with the case of weak screening, when $r\!_s \gg A$.

Consider first chemical potential of small ions. It must be the
same for ions in the bulk solution, ions condensed on the surface
of dissolved DNA rods, and ions on those DNA rods which, in turn,
are adsorbed on the macroion. The two corresponding equations read:
\bea k_BT \ln(N_{1,sb}/N_1) & = & (2e \eta_c / D)\ln(r\!_s/a)
\nonumber
\\ k_BT \ln(N_{1,ss}/N_1) & = & -e \psi(0)+
 e (2 e \eta^{*} /D) \ln (A /2\pi a) \ .
\label{Mono_surf} \eea
The left-hand sides of Eqs. (\ref{Mono_surf}) is obviously the
entropy loss of monovalent ions when they 
move from the solution to the surface of DNA 
and the right-hand side is the potential energy gain
per ion. In the entropic part, $N_{1}$, $N_{1,sb}$ and
$N_{1,ss}$ are the concentrations of counterions in the bulk and
at the surface of the dissolved and adsorbed DNA, respectively.
In the energetic part, $\psi(0)$ is
the average potential of the uniformly over-charged plane with
charge density $\sigma^{*}$, while logarithmic terms correspond to
the potential of DNA having charge density $-\eta_c$ 
in the bulk, where it is screened by monovalent ions
at the distance $r_s$, and DNA with charge density 
$-\eta^{*}$ at the surface, where it is screened
by other DNA molecules at the distance $A/2\pi$. 
As regards $\psi(0)$, it is given by the Gouy-Chapman formula
$\psi(0)\simeq -(2k_BT/e)\ln ( ( r\!_s + \lambda^{*})
/\lambda^{*}) $, where $\lambda^* =Dk_{B}T/ (2\pi e \sigma^*)$,
because plane as a whole is non-linearly screened. 
Concentrations $N_{1,ss}$ and $N_{1,sb}$ are related to the
corresponding net charges: $N_{1,ss}/N_{1,sb} = (\eta_0 -
\eta^{*})/(\eta_0 - \eta_c)$. According to Eq. (\ref{NL2}) 
$\eta^{*}$ differs only logarithmically from $\eta_c$. 
Therefore, we can neglect $\ln (N_{1,ss}/N_{1,sb})$  
and use the same notation $N_{1,s}$ for both concentrations.  
Taking into acount that $\eta_c = k_{B}TD/e$, we get
\beq 
\eta_c \ln(\lambda^*/a) = \eta^{*} \ln(A/2\pi a).
\label{first}
\eeq
Already at this stage we can look at what happens if we neglect
correlations between adsorbed DNA rods. In this case, they are
screened by small ions only, and the corresponding screening
length is $\lambda^{*}$ instead of $A/2\pi$. Eq. (\ref{first})
then gives $\eta^{*}=\eta_c$. Thus, if adsorbed DNA molecules were
uncorrelated, positioned and oriented randomly, they would not
have released counterions upon adsorbtion.

Let us return to the strongly correlated distribution of DNA. 
To complete the picture, we concentrate on the chemical potential
of DNA rods. Corresponding equilibrium condition is similar to Eq.
(\ref{capacitor}):
\bea L\eta^{*}\psi(0) = |\mu_{c}|+ L(\eta^{*}-\eta_c)
(k_BT/e)\ln(N_{1,s}/N_1)
\nonumber \\
- (L\left. \eta^{*} \right.^{2}/D) \ln(\lambda^*/a) +
(L\eta_{c}^{2}/D) \ln(r\!_s/a) \ , \label{DNAmu} \eea
where $L$ is DNA length. In the right hand side, the first term,
$\mu_{c} \simeq \mu_{WC} = - (L\left. \eta^{*}
\right.^{2}/D)\ln(2\pi\lambda^*/A)$, 
is the correlation chemical potential of DNA at the
macroion surface. This is the energy of
interaction of one DNA rod charged to $-\eta^{*}$ with its
correlation hole which is a positive background stripe of width $A$. 
The second term is precisely the entropy gain
due to the release of counterions.  The third and fourth terms
together represent the increase of the DNA self-energy when it
moves from the bulk solution where it is charged to $-\eta_c$ and
screened at $r\!_s$ to the surface where it is charged to
$-\eta^{*}$ and screened at $\lambda^{*}$. 

Fundamental equations (\ref{first}), (\ref{DNAmu})
are identical with Eqs. (57), (60) of Ref~\onlinecite{Shklov0011}, 
which are derived 
there in a more rigorous way. The solution of these equations 
is given by Eqs. (\ref{NL1}), (\ref{NL2}), which simultaneously describe the
correlation induced charge inversion and counterion release.

Let us now continue to explore what happens if we imagine that
DNA is randomly distributed on the macroion surface with respect to both
position and orientation of rods, i. e. if we artificially
drop all correlation effects. As we already
mentioned, Eq. (\ref{first}) yields $\eta^{*} = \eta_c$ in this
approximation. Furthermore, $\mu_{c}=0$ in the absence of
correlations, so that Eq. (\ref{DNAmu}) yields $\psi(0)=0$ and
$\sigma^{*}=0$. Thus, assuming adsorbed DNA randomly distributed
on the macroion surface with respect to both positions and
orientations of rods, we arrive at the conclusion that there is
neither counterion release nor charge inversion. By contrast,
lateral correlations leading to parallel alignment of DNA
molecules, can drive charge inversion even without counterion
release, as this happens, e.g., for weakly charged PE, for
which $\eta_0 < \eta_c$. In this sense, we conclude that
correlations are the driving force of both counterion release and
charge inversion.

Returning to the correlated distribution of DNA 
let us consider also a limit of very large $r\!_s$, 
which is not very realistic, but interesting
from a theoretical standpoint.  
According to the Eq. (\ref{NL2}), DNA net charge $\eta^{*}$
increases with increasing $r\!_s$, which physically means higher
proportion of released counterions.  To address this regime, we
should make one step back including the $\ln (\eta_0
- \eta^{*})/(\eta_0 - \eta_c)$ term in Eq. (\ref{first}). 
For our purposes here it is
sufficient just to approximate $\eta^{*}$ by $\eta_{0}$ in the
Eq. (\ref{DNAmu}).  First and third term in the right-hand side
combine then to $(L\eta_0^2 /D) \ln (A/2\pi a)$, which vanishes at
$A=2\pi a$ and is negligible around this point.  Therefore, the
averaged surface potential in this case
\bea \eta_0 \psi(0) = ( \eta_{0} -
\eta_c) (k_BT/e)\ln(N_{1,s}/N_1) + (\eta_{c}^{2}/D)\ln(r\!_s/a),
  \label{DNAmunew} \eea
looks like the sum of counterion release and metallization terms.
Interestingly, both of them are driven by correlations!

We conclude repeating that the underlying physics of charge
inversion is always determined by correlations.  An explicit
treatment of correlations provides regular and universal
description of this phenomenon.  Apart from charge inversion, the
correlations manifest themselves also in a number of other ways,
including metal-like properties of the macroion surface
over certain range of length scales and including the release of
counterions for the case of highly charged $Z$-ions, such as DNA.

T. T. N. and B. I. S. are supported by NSF DMR-9985985.

\end{multicols}
\end{document}